\documentclass{ifacconf}
\usepackage{graphicx}      % include this line if your document contains figures
\usepackage{natbib}        % required for bibliography

\usepackage{url}            % helped braking the the urls.

\usepackage{customcommands} 

\usepackage{xcolor}
\definecolor{darkgreen}{RGB}{0,128,0}
\usepackage{amsmath,amssymb,amsfonts}
\usepackage{bm}
\usepackage[free-standing-units]{siunitx} % Provides SI units like \meter.
\newcommand{\figref}[1]{Fig.~\ref{#1}}
\newcommand{\tabref}[1]{Tab.~\ref{#1}}
\DeclareSIUnit{\millisecond}{ms}

\usepackage{todonotes}
\usepackage{multirow}
\usepackage{rotating}
\usepackage{booktabs}
\begin{document}
\begin{frontmatter}

%\title{Robust Estimation Strategies For Bluetooth-aided Inertial Navigation\thanksref{footnoteinfo}} 
%\title{Bluetooth-aided multirotor Drone Flight: Offline Experimental Verification\thanksref{footnoteinfo}} 
%\title{Factor Graph-based Bluetooth-Aided Navigation of Multirotor Unmanned Aerial System: Experimental Verification\thanksref{footnoteinfo}} 
%\title{Bluetooth-Inertial Navigation of Multirotor Unmanned Aerial Systems Using Factor Graphs: Experimental Verification\thanksref{footnoteinfo}} 
%\title{Bluetooth-aided Multirotor Unmanned Aerial Systems Navigation Using Factor Graph Optimisation: Experimental Verification\thanksref{footnoteinfo}} 
\title{Bluetooth Phased-array Aided Inertial Navigation Using Factor Graphs: Experimental Verification\thanksref{footnoteinfo}} 

% Title, preferably not more than 10 words.

\thanks[footnoteinfo]{The work is supported by the Research Council of Norway through the project Phased-array radio systems for resilient localization and navigation of autonomous systems in GNSS-denied environments PARNAV (no. 338789). \\\textcopyright\ 2026 the authors. This work has been accepted to IFAC for publication under a Creative Commons Licence CC-BY-NC-ND.}

\author[First]{Glen Hjelmerud Mørkbak Sørensen} 
\author[First]{Torleiv H. Bryne} 
\author[First]{Kristoffer Gryte} 
\author[First]{Tor Arne Johansen} 
%\author[Third]{Third C. Author}

\address[First]{Department of Engineering Cybernetics, Norwegian University of Science and Technology (NTNU) NO-7491 Trondheim, Norway (corresponding author e-mail: glen.h.m.sorensen@ntnu.no).}
%\address[Second]{Colorado State University, 
%   Fort Collins, CO 80523 USA (e-mail: author@lamar. colostate.edu)}
%\address[Third]{Electrical Engineering Department, 
%   Seoul National University, Seoul, Korea, (e-mail: author@snu.ac.kr)}

\begin{abstract}                % Abstract of 50--100 words
Phased-array Bluetooth systems have emerged as a low-cost alternative for performing aided inertial navigation in GNSS-denied use cases such as warehouse logistics, drone landings, and autonomous docking. Basing a navigation system off of commercial-off-the-shelf components may reduce the barrier of entry for phased-array radio navigation systems, albeit at the cost of significantly noisier measurements and relatively short feasible range. In this paper, we compare robust estimation strategies for a factor graph optimisation-based estimator using experimental data collected from multirotor drone flight. We evaluate performance in loss-of-GNSS scenarios when aided by Bluetooth angular measurements, as well as range or barometric pressure.
\end{abstract}

% Five to ten

\end{frontmatter}
%===============================================================================

%%
%% INTRODUCTION
%%
\section{Introduction}

%%%
%%% Background 
%%%
%\subsection{Background}

Global navigation satellite systems (GNSS) are ubiquitous in modern-day aided inertial navigation systems (INS). However, depending on the use case, GNSS may not be available at all (e.g., when indoors) or the availability of GNSS may be severely degraded due to natural or intentional interference (e.g., multipath in a dense urban area or signal jamming or spoofing). Consequently, GNSS-denied INS are necessary in these domains with examples including systems based on LiDAR \cite[]{brossard_associating_2022-1}, vision \cite[]{lupton_visual-inertial-aided_2012}, and phased-array radio systems (PARS) \cite[]{okuhara_phased_2023}. 

Although most PARS-based navigation systems can be categorised as industry- or military-grade, \textit{Bluetooth Low Energy} (BLE) PARS has emerged as a low-cost alternative based on commercial-off-the-shelf (COTS) components for aiding of e.g., fixed-wing UAV flight \cite[]{sollie_automatic_2024}. By basing a navigation system on COTS BLE components, the barrier of entry for employing such phased-array radio systems can be greatly reduced, albeit at the cost of a relatively short operating range and considerably noisier measurements. Furthermore, additional time synchronisation of measurements is necessary as a result of antenna switching and less stable oscillators \cite[]{sollie_outdoor_2022}. Consequently, it is essential to employ robust estimation techniques to handle outliers.

In \cite{sorensen_robust_2025}, we presented an estimation scheme fusing PARS and inertial measurements based on factor graph optimisation (FGO), comparing performance against the industry-standard error-state Kalman filter (ESKF) in a simulation study. Outliers were handled with the natural test (NT), the Huber M-estimator, or the Tukey M-estimator \cite[]{gustafsson_statistical_2010,zhang_parameter_1997} in the presence of simulated sensor faults. In this  paper, we present the following contributions building on our previous publication:
\begin{itemize}
    \item We apply our estimation framework on experimental data from multirotor drone flight, showcasing feasibility of BLE PARS-aided INS.
    \item We compare the performance of our FGO-based estimator on the \SE{3} matrix Lie group against a benchmark ESKF in handover scenarios where the drone goes from using Real-time kinematics (RTK)-GNSS aiding with position and compass measurements to (1) BLE PARS and RTK-GNSS range and (2) BLE PARS and barometric pressure. 
   % \item We conduct a more rigurous tuning process of M-estimators and smoother lag by evaluating performance when simulated measurements faults are applied to the experimental data.
   % \item We provide the source code in a public GitHub repository.
\end{itemize}

RTK-based range measurements are used as an intermediary step since it allows us to experimentally verify the methods in the face of e.g., multipath. Integration of Bluetooth LE range from channel sounding \cite[]{bt_cs_overview} is planned.

%\todo[inline]{Nevne hvorfor vi bruker RTK range som et mellomsteg?}

%%
%% PRELIMINARIES
%%
\section{Preliminaries}

\subsection{Notation and coordinate frames}
%\todo[inline]{Get cleveref working, doesn't seem to want to work with this template}
\iffalse
An overview of the notation used throughout this paper is found in \tabref{tab:notation}
\begin{table}[tb]
\label{tab:notation}
\centering
\caption{Overview of notation.}
\begin{tabular}{cc}
Notation    & Description \\ \hline
\bvec{a}, \bvec{A}    & Vector, matrix            \\
\nedframe{}    & North-East-Down (NED) frame            \\
\bodyframe{}{} / \radioframe{} / \sensorframe{}  & Body / Radio / Sensor frame            \\
\rotmat{b}{n} & Rotation matrix between \bodyframe{}{} and \nedframe{} \\     
\liepose{b}{n} & Pose between \bodyframe{}{} and \nedframe{}  \\
$\bvec{a}^{n}$ & Vector decomposed in \nedframe{}\\ 
$\hat{\bm{x}}$ & An estimate of state $\bvec{x}$ \\ \hline
\end{tabular}
\end{table}
\fi

Vectors and matrices are given in bold face, cursive lowercase \bvec{v} and uppercase letters \bvec{A}, respectively. \rotmat{a}{b} represents the rotation matrix between two coordinate frames, i.e., from frame $\{a\}$ to frame $\{b\}$. In this paper, four frames are considered: local North-East-Down navigation frame \nedframe{}, the BODY-frame \bodyframe{}{}, the inertial measurement unit (IMU) sensor frame \sensorframe{}, and the Bluetooth PARS radio frame \radioframe{}. E.g., $\pos{r}{b}{n}$ denotes the position measured in \bodyframe{}{} relative to \radioframe{}, decomposed in \nedframe{}. Estimates are expressed with a hat, e.g., $\hat{x}$ is an estimate of $x$.

%\nedframe{} denotes the North-East-Down local navigation frame, \bodyframe{} denotes the BODY-frame, \sensorframe{} denotes the IMU sensor frame, and \radioframe{} denotes the Bluetooth radio frame. \rotmat{b}{n} denotes the rotation matrix from the BODY frame \bodyframe{} to the North-East-Down local navigation frame \nedframe{}.

\subsection{SE(3) matrix Lie group theory}

The \SE{3} matrix Lie group is defined as the set of poses $\liepose{}{} \in \realdomain{4\times 4}$ 
\begin{equation}
    \SE{3} \triangleq{} 
    \left\{ 
    \left.
    \transmat{}{}  =  
    \begin{bmatrix}
        \rotmat{}{} & \pos{}{}{} \\
        \bm 0_{1\times 3} & 1 
    \end{bmatrix}
    \right| 
    \begin{array}{ll}
         \rotmat{}{} \in \SO{3},\\
         \pos{}{}{} \in \realdomain{3}
    \end{array}
    \right\},
\end{equation}
with the corresponding \textit{Lie algebra} \se{3} given by the set of matrices
\begin{equation}
    \se{3} \triangleq{} 
    \left\{ 
    \left.
    \bm\xi^\wedge{}  =  
    \begin{bmatrix}
        \skewmat{\xirot{}} & \xitrans{}\\
        \bm 0_{1\times 3} & 0 
    \end{bmatrix}
    \right| 
    \begin{array}{ll}
         \skewmat{\xirot{}} \in \so{3},\\
         \xitrans{} \in \realdomain{3}
    \end{array}
    \right\},
\end{equation}
where $\bm{\xi}_\bullet$ is a small, local perturbation mapped to the Lie algebra using the hat operator $^\wedge$. The \textit{exponential map} of the Lie group is subsequently defined using the matrix exponential: $\Exp{\bm{\xi}} \triangleq{} \exp \left ( \bm{\xi}^\wedge{} \right )$, and maps a matrix in the Lie algebra onto the Lie group itself. The reader is referred to \cite{barfoot_associating_2014} for details.

\section{Bluetooth LE direction finding}

The angle-of-arrival (AoA) measurements from the Bluetooth PARS receiver are obtained using direction finding. This involves appending a \textit{constant tone extension }(CTE) to each advertisement packet sent from the transmitter, which is subsequently sampled by the antennae at the receiver. The raw samples are then transformed into angular measurements (azimuth $\Psi$ and elevation $\alpha$), similar to the ones found in \cite{gryte_field_2019}:
\begin{equation}\label{eq:locator-frame-meas-model}
\begin{split}
    \obsv{}_\text{PARS}^{r} &= 
    \begin{bmatrix}
     %   \rho^{r} \\
        \Psi^{r} \\
        \alpha^{r}
    \end{bmatrix} =
    \begin{bmatrix}
      %  \| \pos{r}{b}{r} \|_2 \\
        %\arctan \big( \frac{p^r_{rb,y}}{p^r_{rb,x}} \big) \\
        \arctanTwo{p^r_{rb,y}}{p^r_{rb,x}} \\
        %\arctanTwo{p^r_{rb,y}, {p^r_{rb,x}} } \\
        %\arctan \big(  -\frac{p^r_{rb,z}}{\bar \rho} \big)
        \arctanTwo{-p^r_{rb,z}}{\bar \rho}
    \end{bmatrix} + \bvec{\varepsilon}, \\
    %&= \bvec{h}_\text{PARS}(\bvec{x})+\bvec{\varepsilon},
\end{split}
\end{equation}
where $\bvec{\varepsilon}$ is zero-mean Gaussian noise and $\bar \rho$ is the horizontal range of the measured position in \radioframe{}, i.e., the Euclidean distance between the horizontal components and the origin of \radioframe{}. This is the same model we used in \cite{sorensen_robust_2025}, with the notable exclusion of the range, which is not available with direction finding. The rotation matrix \rotmat{r}{n} relating \radioframe{} to \nedframe{} needs to be estimated via rough calibration based on mounting or through some form of calibration algorithm using GNSS, as done in e.g., \cite{okuhara_phased_2023}. If the origin of \radioframe{} differs from that of \nedframe{}, the lever arm $\bm{l}^n_{PARS}$ must also be accounted for, i.e, $\pos{r}{b}{n} = \rotmat{r}{n} \pos{r}{b}{r} +\bm{l}^n_{PARS}$. The reader is referred to \cite{df_bt_wpf} for details on the underlying direction finding technology and to \cite{sollie_outdoor_2022} for details on the algorithm used for obtaining the angular measurements, as both are outside the scope of this paper.

\section{Inertial Navigation system}

This section presents the components of the FGO-based aided inertial navigation system. At its core, it is the same navigation system presented in \cite{sorensen_robust_2025}, which is based on the iSAM2 fixed-lag smoother from the GTSAM C++ library developed by \cite{dellaert_borglabgtsam_2022}. 
%\todo[inline]{Taken form ECC paper. Rerwite}

The estimator maintains separate state- and covariance estimates of the pose, velocity, and IMU biases, i.e., $\hat{\bm{x}}\triangleq{}(\transmathat{b}{n}, \velhat{n}{b}{n},\hat{\bm{b}}^b ) \in \SE{3} \times \realdomain{3} \times \realdomain{6}$, where $\hat{\bm{b}}^b\triangleq{}[\accbiashat{b,\top}~\gyrobiashat{b,\top}]^\top$, The estimation itself is based on minimising 
\begin{align}
    \transmathat{}{} &= 
    \underset{ \transmat{}{} }{\text{arg min}}
    \| \bm h(\transmat{}{}) - \bm z\|^2_{\bm{\mathcal{P}}},
\end{align} 
in the pose optimisation given in Ch. 6.2 in \cite{dellaert_factor_2017} (and in equivalent expressions for the velocity and bias states). To solve the problem with respect to the local perturbation $\bm \xi$, the problem is reformulated to
\begin{align}    %
    \bvec{\hat{\xi}}
    &\approx
    \underset{ \bm \xi }{\text{arg min}}
    \| \bm h(\transmathat{}{}) + \bm H \bm \xi - \bm z\|^2_{\bm{\mathcal{P}}}
    \label{eq:pose_opt_form_gtsam},
\end{align}
where $\bm H$ is the measurement Jacobian with respect to \se{3} (or velocity or bias) and $\bm{\mathcal{P}}$ is the given estimation error covariance. % Incorporation of velocity and biases can be handled by augmenting $\transmat{}{}$ such that
%$\transmat{\mathrm{aug}}{} = \blkdiag(
%\transmat{}{},
%\transmat{\mathrm{vel}}{},
%\transmat{\mathrm{ars}}{},
%\transmat{\mathrm{acc}}{})$.
We first present the different measurement factors, before presenting the core navigation system. 

\subsection{On-manifold IMU preintegration factor}
For on-manifold IMU preintegration, we use the model presented in \cite{forster_imu_2015} to encode IMU measurements between two time steps $i$ and $j$ in an efficient manner:
\begin{equation}\label{eq:preint-model}
    \begin{split}
        \rotpreint{} &= \rotmat{i}{\top} \rotmat{j}{} \Exp{\processnoiseattitude{}{}} \\
        \velpreint{} &= \rotmat{i}{\top} (\vel{j}{}{} - \vel{i}{}{} - \gravity{} \timepreint{}) + \processnoisevelocity{}{}\\
        \pospreint{} &= \rotmat{i}{\top} (\pos{j}{}{} - \pos{i}{}{} - \vel{i}{}{}\timepreint{} - \frac{1}{2} \gravity{} \timepreint{}^2 ) + \processnoiseposition{}{}.
    \end{split}
\end{equation}
Here, $\anypreint{}$ denotes the \textit{preintegrated measurement} of the given state between the two time steps, $\timepreint{}$ is the time between the two given time steps, $\gravity{}$ is the gravity vector, and $\bm{w}_\bullet$ is zero-mean Gaussian noise for angular velocity, acceleration, and the preintegration itself (from top to bottom). The implementation we use is GTSAM's internal on-manifold IMU preintegration factor, which is developed based on the above reference, as well as \cite{lupton_visual-inertial-aided_2012,carlone_eliminating_2014}.

The pose and velocity estimates are propagated in between measurement updates using \eqref{eq:preint-model}, whilst the bias estimates are updated whenever an optimisation is performed on the graph.

\subsection{GNSS position and attitude factors}
As a reference we employ the GNSS Factor (GPSFactor internally in GTSAM)
\begin{subequations}\label{eq:gnss-jacobian}
%\vspace{-0.5cm}
\begin{alignat}{3}
    \obsv{}_\text{RTK}^{n} &= \pos{n}{b}{n} + \bvec{\varepsilon}, \\
    \bm h_\text{RTK}(\transmathat{}{}) &= \poshat{n}{b}{n}, \\
    \jacobian{\text{RTK}} &= \left[\OO{3\times3}~\rotmathat{b}{n}\right], \label{eq:H_p}
\end{alignat}
\end{subequations}
and a factor for GNSS compass
\begin{subequations}\label{eq:gnss-att-jacobian}
%\vspace{-0.5cm}
\begin{alignat}{3}
    \obsv{}_\text{comp}^{n} &= \rotmat{b}{n} \bm{l}_\text{RTK}^{b} + \bvec{\varepsilon}, \\
    \bm h_\text{comp}(\transmathat{}{}) &= \rotmathat{b}{n} \bm{l}_\text{RTK}^{b}, \\
    \jacobian{\text{comp}} &= \left[-\rotmathat{b}{n} \skewmat{\bm{l}_\text{RTK}^{b}} ~\OO{3\times3}  \right], 
\end{alignat}
\end{subequations}
where $\bm{l}_\text{RTK}^{b}$ is the RTK baseline between the two GNSS antennas and $\bm{\varepsilon}$ is zero-mean Gaussian noise. These two factors are used before the handover to Bluetooth PARS-based aiding measurements to the navigation system. 
\subsection{Bluetooth factors}
The Jacobians of the Bluetooth factors on \SE{3} are based on \eqref{eq:locator-frame-meas-model}. Since BLE range is not present in the experimental data, we utilise range derived from the relative NED position from RTK in the interim
\begin{equation}\label{eq:rtk-range}
\begin{split}
    \obsvscalar{}{}_\text{RTK\,range}^{r} &= \rho^{r}  = \| \left ( \rotmathat{r}{n} \right)^\top \left ( \pos{n}{b}{n} - \bm{l}^n_{PARS} \right ) + \bvec{\varepsilon} \|_2 , \\
\end{split}
\end{equation}
and treat this as BLE PARS range as presented in our previous work. Thus, the Jacobians are given by
\begin{subequations}\label{eq:pars-jacobians}
%\vspace{-0.5cm}
\begin{alignat}{3}
\jacobian{\rho} &= \frac{1}{\| \poshat{r}{b}{r} \|_2} (\poshat{r}{b}{r})^\top \jacobian{p}\\
\jacobian{\Psi} &= \frac{ 1}{\hat p_x^2 + \hat p_y^2} 
        \begin{bmatrix}
            -\hat p_y & \hat p_x & 0
        \end{bmatrix}\jacobian{p}\\
\jacobian{\alpha} &= 
           \frac{1}{\| \poshat{r}{b}{r} \|_2^2} 
         \begin{bmatrix}
            \frac{p_x p_z}{\sqrt{\hat{p}_x^2 + \hat{p}_y^2}} & \frac{p_y p_z}{\sqrt{\hat{p}_x^2 + \hat{p}_y^2}} & -\sqrt{\hat{p}_x^2 + \hat{p}_y^2}
        \end{bmatrix}\jacobian{p},
\end{alignat}
\end{subequations}
where $\jacobian{p} = [\OO{3\times3}~( \rotmathat{r}{n} )^\top \rotmathat{b}{n}]$ relates the position measured by the locator to \se{3}. The Jacobians and corresponding factors were derived in \cite{sorensen_robust_2025} and we refer to that paper for details. The difference between the two setups is that \pos{r}{b}{n} is replaced with $\rotmathat{r}{n}\pos{r}{b}{r}$ in the derivations to get the Jacobians on \radioframe{}.

\subsection{Barometric factor}
We also use a barometer to provide vertical corrections with a factor based on GTSAM's internal barometric factor which is implemented based on the Earth atmosphere model found online \cite[]{nasa_glenn_research_center_earth_2013}:
%\iffalse
\begin{subequations}
    \begin{alignat}{3}
    \obsvscalar{}_{\mathrm{baro}}
     &= 
    \frac{\frac{p_\mathrm{baro,comp}}{101.29}^{1/5.256} \cdot 288.08 -288.14}{-0.00649} + \varepsilon{}
     \\
    p_\mathrm{baro,comp} &= p_\mathrm{baro} - b_\mathrm{baro}  \\
    \hat{z}&= 
    \begin{bmatrix}
            0 & 0 & -1
    \end{bmatrix}\poshat{n}{b}{n}%  + \hat{b}_\mathrm{baro}
    \\
    \bm H_\mathrm{baro,\SE{3}} &= 
    \begin{bmatrix}
        \bm 0_{1 \times 3}  & 
        \begin{bmatrix}
            0 & 0 & 1
        \end{bmatrix}\rotmathat{b}{n}
    \end{bmatrix} 
\end{alignat}
\end{subequations}
where $\varepsilon{}$ is zero-mean Gaussian noise and $p_\mathrm{baro}$ is the barometric pressure measurement given in kPa. We calculate the pressure bias and statically compensate for $b_\mathrm{baro}$ before take-off directly to the pressure measurement.

\subsection{Propagating the estimate and dealing with outliers}

The INS uses the predicted state from the preintegrator between measurement updates. Whenever an aiding measurement is available, a factor representing the preintegrated IMU measurement since the last update is added to the factor graph alongside the aiding measurement factor(s). The fixed-lag smoother then calculates the updated estimate and resets the preintegrator with the updated IMU bias estimate.

Each factor will have an uncertainty associated with it, typically modelled as zero-mean Gaussian noise. In the implementation, the noise model is encapsulated within a robust loss function representing one of two \textit{M-estimators} in order to mitigate outliers. Originally proposed in \cite{huber_robust_1964}, M-estimators are a group of estimators suitable to mitigating the effect of outliers without necessarily rejecting them outright. Here, they differ from the natural test \cite[Ch.~7.6]{gustafsson_statistical_2010}, which is a technique that rejects a measurement outright if the test statistic exceeds the given threshold. 

In this study, we applied the \textit{Tukey} and \textit{Geman-McClure} M-estimators to the experimental multirotor data. They are described by their objective/cost functions $Q_\bullet(\tilde{z})$
\begin{align}
%\begin{equation}
\label{eq:tukey}
     Q_\mathrm{T}(\tilde{z})&= 
   \begin{cases}
        \frac{c^2}{6}\left( 1 - [1 - (\tilde{z} - (\tilde{z}/c)^2]^3 \right), & \text{for }|\tilde{z} | \leq  c \\
       \frac{c^2}{6}, & \text{otherwise,}
    \end{cases}  \\
%\end{equation}
%\begin{equation}
  %  Q_\mathrm{G}(\tilde{z}) &= \frac{\tilde{z}^2}{c^2 + \tilde{z}^2},
     Q_\mathrm{G}(\tilde{z}) &= \frac{1}{2}\frac{c^2 \tilde{z}^2}{c^2 + \tilde{z}^2},
    \label{eq:gmc}
%\end{equation}
\end{align}

for Tukey (TK) and Geman-McClure (GM), respectively. $c$ is the given estimator bound and $\tilde{z}$ denotes the residual between the measurement and its predicted value (\textit{innovation}). We note that the two estimators behave differently – Tukey caps large outliers at a constant max value if the residual exceeds the threshold, whereas Geman-McClure smooths the curve in a continuous fashion, suitable for real-time use. In our previous paper, we considered the Huber M-estimator alongside Tukey and found that the former's inclusion of outliers no matter how large resulted in poorer performance, which is why we omit it from this study. The reader is referred to \cite{zhang_parameter_1997,sh-ch3-outlier} for further details on the M-estimators.

%Here, the left-hand side represents \emph{compound measurements} obtained by integrating the IMU measurements between aiding measurements. Since it is the specific force vector measured by the accelerometers that is integrated, $\velpreint{}$ and $\pospreint{}$ do not represent true increments, although $\rotpreint{}$ does. 
%To represent the measurements from the drone's IMU, we use GTSAM's internal on-manifold IMU preintegration factor, which is developed based on \cite{lupton_visual-inertial-aided_2012,carlone_eliminating_2014,forster_imu_2015}. 

%\subsection{On-manifold IMU preintegration}

%\todo[inline]{Zhang presenterer kun GMC uten skalering}

%%
%% EXPERIMENT SETUP
%%
\section{Experiment setup}

%This section presents the setup of the multirotor drone experiment used for experimental verification.

%\subsection{Hardware, measurements, and drone trajectory}

\begin{figure}[tb]
    \centering
    %\missingfigure[]{test}
    \includegraphics[width=0.65\linewidth]{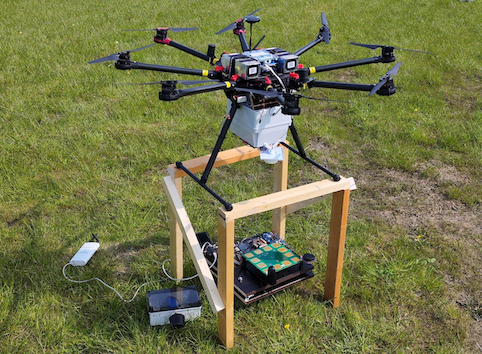}
    \vspace*{-0.25cm}
    \caption{Multirotor drone and ground station in the field. The Bluetooth antenna array (green) is mounted flat.}
    \label{fig:drone-groundstation}
\end{figure}

\begin{figure}[tb]
    \centering
    \includegraphics[width=0.90\linewidth]{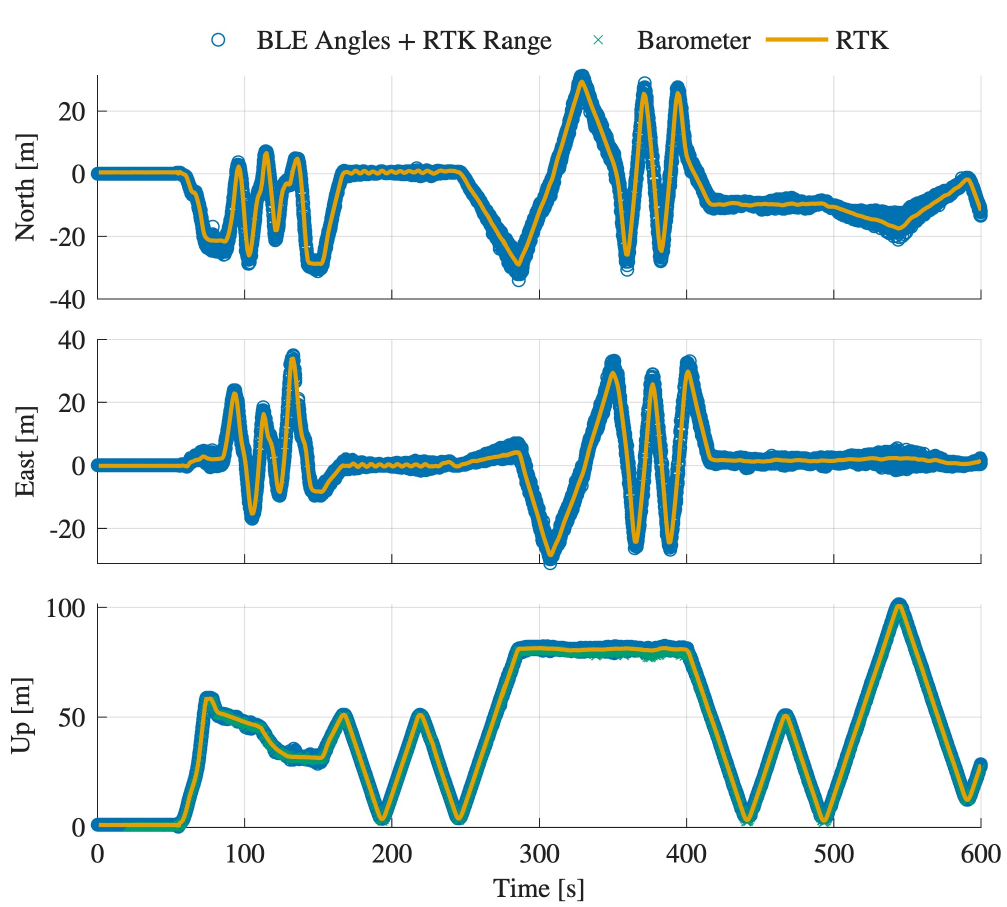}
    \vspace*{-0.25cm}
    \caption{NED position computed from measurements.}
    \label{fig:autopilot-path}
\end{figure}

The multirotor drone flight experiment data were obtained from an experimental campaign at one of NTNU's test sites in a field west of Trondheim. The drone and ground station are both equipped with a Rasperry Pi 4B, Nordic Semiconductor nRF52833 BLE devkit, and Sentisystems Sentiboard v1.3. The ground station BLE devkit is connected to a 12-element antenna array, whilst the drone is equipped with a STIM300 IMU, a barometric pressure sensor, and a uBlox dual RTK GNSS receiver. The drone runs a modified version of the DUNE robotic middleware \cite[]{pinto_lsts_2013} and uses the Pixhawk Cube Orange autopilot with ArduCopter v4.6.0.  %, shown in \figref{fig:groundstation}. 
%\end{itemize}
The drone and ground station are shown in \figref{fig:drone-groundstation}. After takeoff, the wooden frame was moved as to not obstruct line-of-sight between the drone-mounted transmitter and ground station receiver. During the experiment, the drone followed a preconfigured path given by the autopilot software. The relative NED position computed from the measurements is shown in \figref{fig:autopilot-path}. %The altitude as measured by the RTK and via the barometric pressure are shown in \figref{fig:baro}. 

The empirical update rates of the sensors used are approximately: 2000Hz for the IMU, 1Hz for RTK, 2Hz for barometer, and 16.6Hz for BLE PARS. In order to obtain range measurements for PARS, the RTK position measurements are interpolated at the IMU frequency and measurements that match BLE PARS timestamps are used to compute range for that given time. % The tuning parameters for the noise were derived from knowledge of the given sensors, see the Github repo \cite[]{sorensen_gsorensenparnav_ins_simulator_nodate} for details. 
The following pre-processing is performed:
\begin{itemize}
    \item IMU specific force and angular velocity measurements are transformed from \sensorframe{} to \bodyframe{}{} using \rotmat{s}{b}, which is roughly known.
    \item The raw IQ samples from Bluetooth are converted into angular measurements.% and transformed into \nedframe{} using \rotmat{r}{n} and $\bm{l}^n_{PARS}$ obtained from rough calibration of the mounting angles and lever arm. 
    \item Using time-aligned RTK and BLE observations, estimates of  \rotmat{r}{n} and $\bm{l}^n_{PARS}$ are found using the FOAM algorithm (\cite{markley_attitude_1993}).
\end{itemize}

%\subsection{Autopilot trajectory}

%\subsection{Estimator configurations and evaluation criteria}

We compare the FGO-based estimation scheme with smoother lag set to $t_{\text{lag}}=\num{5}{\second}$ against a benchmark ESKF in a handover scenario, where the drone switches from navigating aided by RTK position and compass measurements to aiding by Bluetooth angular measurements and either RTK-based range or barometric pressure measurements. In the latter case, the barometer measurements are used before handover to allow the barometer bias estimate to converge.

Outside of the nominal comparison (i.e., using BLE PARS measurements without any outlier rejection or mitigation) the following configurations are tested:
\begin{itemize}
    \item ESKF + Outlier rejection with the NT with $k = 3.841$ 
    \item FGO + Tukey M-estimator with $c = 3.6851$
    \item FGO + Geman-McClure M-estimator with $c=1$
\end{itemize}
The tuning of the different methods are kept as close to one another as possible. The interpolated position from RTK in \nedframe{} at the frequency of the IMU is chosen as ground truth for position. We compare the roll and pitch estimates with the autopilot's own estimates, whilst the yaw computed from the dual GNSS receiver is chosen as yaw reference. We evaluate performance using the root mean square error (RMSE) of each substate, as well as the estimation error and corresponding $3\sigma$-bounds. 

We use the QUEST algorithm to obtain our initial estimate of the drone attitude whilst stationary, see \cite{shuster_three-axis_1981} for details. The GNSS compass is used for further stabilisation of the attitude before the handover, and is not used afterwards.

%%
%% RESULTS AND DISCUSSION
%%
\section{Results and discussion}

\begin{figure}[tb]
    \centering
    \includegraphics[width=0.90\linewidth]{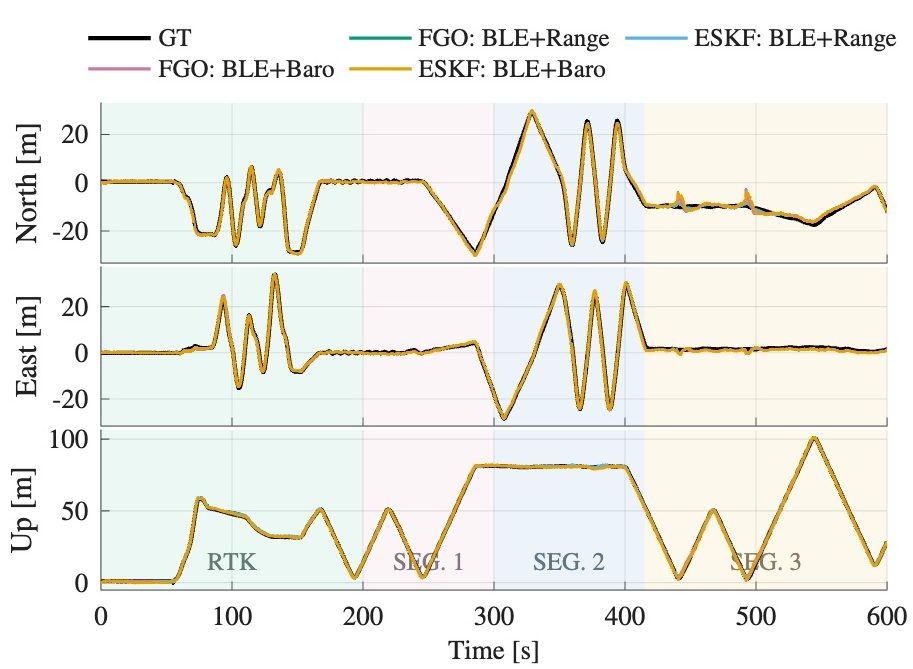}
    \includegraphics[width=0.90\linewidth, height=140px]{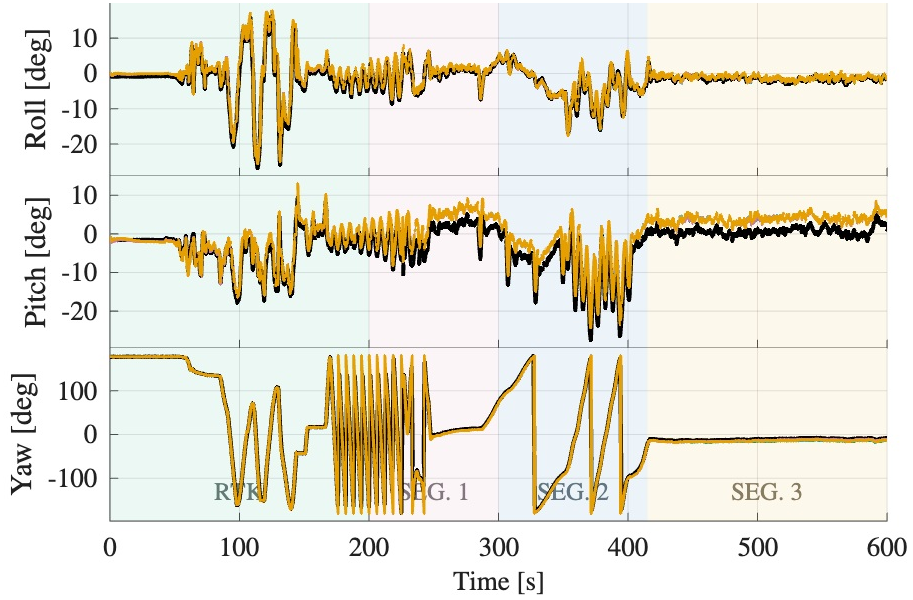}
%    \missingfigure[]{Estimation error RTK attitude}
    \vspace*{-0.3cm}
    \caption{Estimator performance without mitigation of BLE PARS outliers. Note the shaded sections roughly segmenting the data based on RTK/PARS and the manoeuvre performed during the given time interval.}
    %\caption{Estimator performance when BLE PARS measurements are used as is. Data is segmented between RTK and alternative aiding, and later on the manoeuvre performed during the given time interval.}
    \label{fig:est_err_rtk}
\end{figure}

This section presents the performance of the ESKF and FGO-based estimators and discusses the findings. For comparison, the position and attitude estimates when the BLE measurements are used without any outlier mitigation/rejection are shown in \figref{fig:est_err_rtk}. The data is roughly segmented  based on the manoeuvre and whether we have switched from RTK, and we will refer to these segments as \textit{segments} throughout this section. We highlight two problematic areas – one shortly after the set handover point in section 1 and most of section 3.

\subsection{BLE angular measurements and RTK range}

\begin{figure}[tb]
    \centering
    \includegraphics[width=0.90\linewidth]{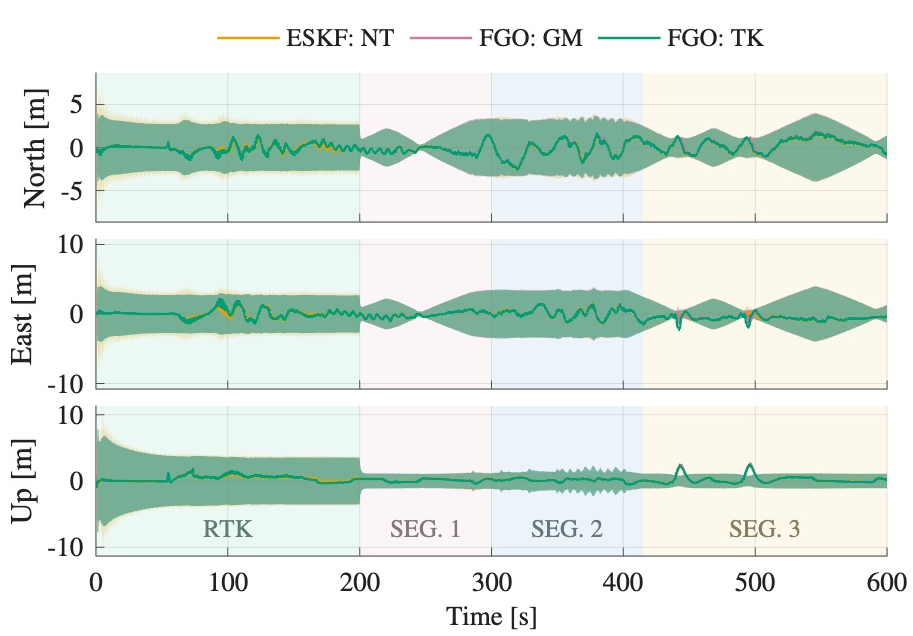}
    \includegraphics[width=0.90\linewidth]{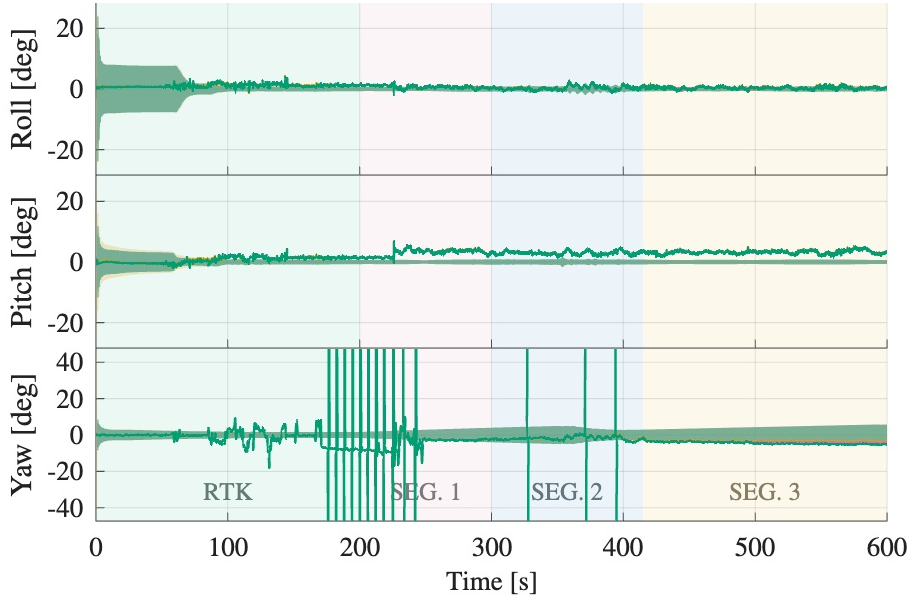}
    \vspace*{-0.30cm}
   % \caption{Estimation error in handover from RTK to BLE angular measurements and RTK range. The respective $3\sigma$-bounds are shaded in the same colour.}
    \caption{Estimation error and $3\sigma$-bounds in handover from RTK to BLE angular and RTK range.}
    %\caption{Estimation error in handover from RTK to BLE angular measurements and RTK range. $3\sigma$-bounds are given by dashed lines of the same colour.}
    \label{fig:error_plots_angle_range}
\end{figure}

The estimation error for NED position and attitude in the handover case to BLE angular measurements and RTK range are shown in \tabref{tab:angle_range_rmse} and \figref{fig:error_plots_angle_range}. We observe comparable performance across the board for position. There are a couple of spikes in Down-position in section 3, and we generally observe good estimation of the NE-error, albeit with a couple of more prominent spikes in East-position error towards the end for FGO with Tukey. Given that this is one of the problematic areas, it may be that Tukey here has weighted these measurements more than the ESKF with the natural test or FGO with Geman-McClure, given the former has a binary cutoff point and the latter is far more aggressive in its de-weighting.

The rapid yaw changes in segment 1 are challenging for all configurations, resulting in several spikes in estimation error  wrapping around $\pm$\num{180}{\degree}, resulting in a large RMSE. However, we note that even before RTK is lost, the estimator is struggling with this manoeuvre, which may indicate that e.g., the gyroscope bias is not properly estimated. By incorporating information from, e.g., magnetometer aiding or using multiple PARS antennas on the drone, this issue may be mitigated. The roll angle is well-estimated, but all estimators appear overconfident.  For pitch, we observe a bias between the autopilot's and our estimate after the handover. A possible explanation may be misalignment in specified mounting angles of the PARS antenna array or a delay/time synchronisation issue in the AoA measurements or artefacts from the AoA estimation. We observe the same tight $3\sigma$-bounds as for roll (and yaw), which may indicate that the IMU measurement noise parameters are set lower than they should be, given the IMU is the only source of info for attitude after handover. %As seen in \figref{fig:autopilot-path}, the RTK and BLE-computed trajectories do not match perfectly throughout the entire run, so it is not unreasonable to assume this may have had an effect on performance.
%\todo[inline]{AoA delay, artefacts from AoA estiamtion, misalignments} 

\begin{table}[tb]
    \centering
    \caption{RMSE BLE angular + RTK range.}
    \label{tab:angle_range_rmse}
    \setlength{\tabcolsep}{5pt}
    \begin{tabular}{c l c c c c c c}
        \toprule
        &         &N [m] & E [m] & D [m] & $\phi$ [$^{\circ}$] & $\theta$ [$^{\circ}$] & $\psi$ [$^{\circ}$] \\
        \midrule
        % RTK
        \multirow{3}{*}{\begin{sideways}RTK\end{sideways}}
            & NT & \textbf{0.399} & \textbf{0.486} & \textbf{0.499} & 1.175 & 1.206 & 23.185  \\
            & TK & 0.458 & 0.578 & 0.575 & \textbf{1.165} & \textbf{1.186} & \textbf{23.168}  \\
            & GM & 0.458 & 0.578 & 0.575 & \textbf{1.165} & \textbf{1.186} & \textbf{23.168}  \\
        \midrule
        % SEG. 1
        \multirow{3}{*}{\begin{sideways}SEG. 1\end{sideways}}
            & NT & 0.566 & \textbf{0.389} & 0.267 & 0.866 & 3.231 & 41.935  \\
            & TK & 0.567 & 0.395 & 0.248 & \textbf{0.855} & \textbf{3.197} & \textbf{41.907}  \\
            & GM & \textbf{0.552} & 0.398 & \textbf{0.247} & \textbf{0.855} & \textbf{3.197} & \textbf{41.907}  \\
        \midrule
        % SEG. 2
        \multirow{3}{*}{\begin{sideways}SEG. 2\end{sideways}}
            & NT & \textbf{1.066} & 0.719 & 0.276 & 0.718 & 3.045 & 19.035  \\
            & TK & 1.068 & 0.682 & \textbf{0.267} & \textbf{0.702} & 3.029 & \textbf{18.525}  \\
            & GM & 1.076 & \textbf{0.664} & 0.271 & 0.703 & \textbf{3.027} & 18.584  \\
        \midrule
        % SEG. 3
        \multirow{3}{*}{\begin{sideways}SEG. 3\end{sideways}}
            & NT & \textbf{0.753} & \textbf{0.635} & 0.619 & 0.560 & 3.331 & \textbf{3.414}  \\
            & TK & 0.775 & 0.700 & \textbf{0.597} & 0.521 & 3.267 & 4.468  \\
            & GM & 0.812 & 0.643 & 0.618 & \textbf{0.516} & \textbf{3.256} & 3.874  \\
        \bottomrule
    \end{tabular}
\end{table}
\begin{table}[tb]
    \centering
    \caption{RMSE BLE angular + barometer.}
    \label{tab:angle_baro_rmse}
    \setlength{\tabcolsep}{5pt}
    \begin{tabular}{c l c c c c c c}
        \toprule
        &         &N [m] & E [m] & D [m] & $\phi$ [$^{\circ}$] & $\theta$ [$^{\circ}$] & $\psi$ [$^{\circ}$] \\
        \midrule
        % RTK
        \multirow{3}{*}{\begin{sideways}RTK\end{sideways}}
            & NT & \textbf{0.399} & \textbf{0.487} & 0.392 & 1.170 & 1.207 & 23.185  \\
            & TK & 0.455 & 0.578 & \textbf{0.378} & \textbf{1.161} & \textbf{1.185} & \textbf{23.168}  \\
            & GM & 0.455 & 0.578 & \textbf{0.378} & \textbf{1.161} & \textbf{1.185} & \textbf{23.168}  \\
        \midrule
        % SEG. 1
        \multirow{3}{*}{\begin{sideways}SEG. 1\end{sideways}}
            & NT & 0.657 & \textbf{0.401} & \textbf{0.332} & 0.869 & 3.233 & 41.938  \\
            & TK & 0.632 & 0.404 & 0.337 & \textbf{0.857} & \textbf{3.195} & \textbf{41.907}  \\
            & GM & \textbf{0.616} & 0.407 & 0.336 & \textbf{0.857} & \textbf{3.195} & \textbf{41.907}  \\
        \midrule
        % SEG. 2
        \multirow{3}{*}{\begin{sideways}SEG. 2\end{sideways}}
            & NT & \textbf{1.170} & 0.808 & 0.563 & 0.727 & 3.040 & 18.956  \\
            & TK & 1.178 & 0.771 & 0.567 & \textbf{0.711} & 3.022 & \textbf{18.491}  \\
            & GM & 1.193 & \textbf{0.760} & \textbf{0.561} & \textbf{0.711} & \textbf{3.020} & 18.549  \\
        \midrule
        % SEG. 3
        \multirow{3}{*}{\begin{sideways}SEG. 3\end{sideways}}
            & NT & \textbf{1.133} & \textbf{0.643} & 1.089 & 0.561 & 3.317 & 3.453  \\
            & TK & 1.139 & 0.722 & 0.970 & 0.533 & 3.239 & \textbf{3.044}  \\
            & GM & 1.151 & 0.753 & \textbf{0.887} & \textbf{0.523} & \textbf{3.230} & 3.327  \\
        \bottomrule
    \end{tabular}
\end{table}

\subsection{BLE angular measurements and barometric pressure}

\begin{figure}[tb]
    \centering
    \includegraphics[width=0.90\linewidth]{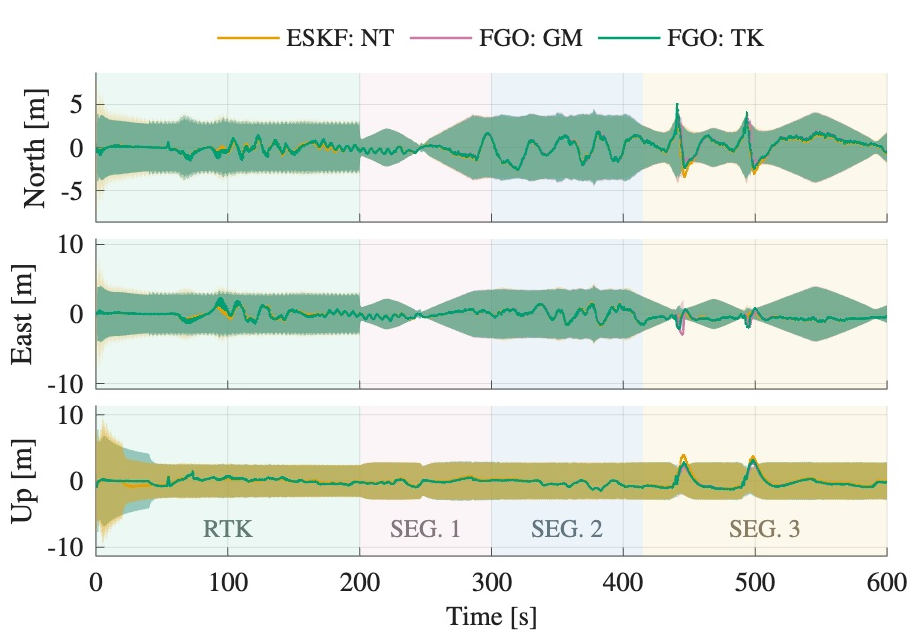}
    \includegraphics[width=0.90\linewidth]{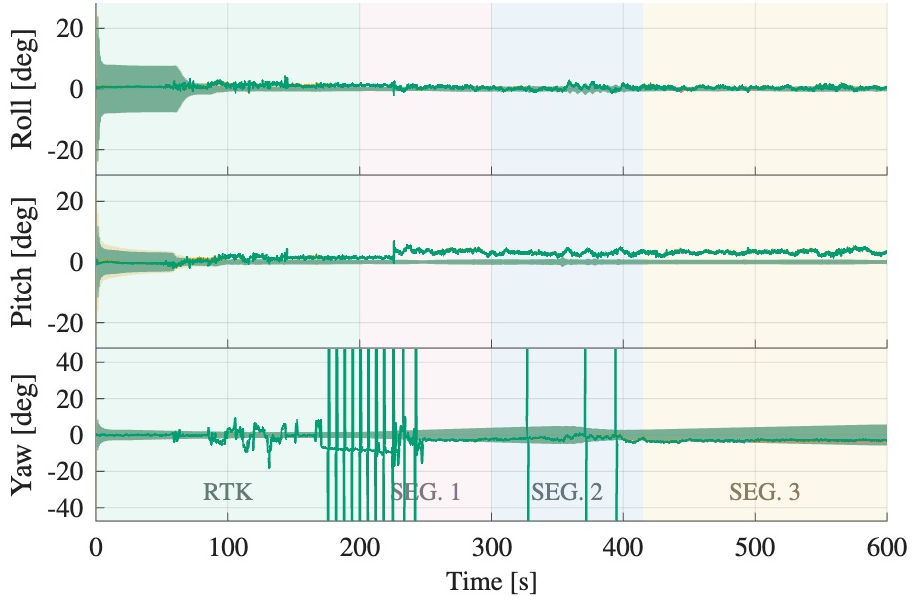}
    \vspace*{-0.30cm}
    %\caption{Estimation error in handover from RTK to BLE angular measurements and barometric pressure. The respective $3\sigma$-bounds are shaded in the same colour.}
   \caption{Estimation error and $3\sigma$-bounds in handover from RTK to BLE angular and barometric pressure.}
    %\caption{Estimation error in handover from RTK to BLE angular measurements and RTK range. $3\sigma$-bounds are given by dashed lines of the same colour.}
    \label{fig:error_plots_angle_baro}
\end{figure}

The estimation results for position and attitude in the handover case to BLE angular measurements and barometric pressure are shown in \tabref{tab:angle_baro_rmse} and \figref{fig:error_plots_angle_baro}. Performance is generally comparable to the first case, but we observe slightly greater position uncertainty after handover, with the ESKF affected the most. This can also be seen in the relevant RMSE values. The Down-position estimate in the problematic area of section 3 deteriorates a fair bit, possibly due to the fewer barometer measurements available coupled with increased rejection/de-weighting of BLE angular measurements. Furthermore, the updates from baro and BLE will in general not be synchronised like those in the previous case. Together, this may explain some of the more notable fluctuations in the estimation error in this case.
%
%\subsection{BLE angular measurements only}
%%
%% CONCLUSION
%%
\section{Conclusion}
%\todo[inline]{Three sentences about application and future avenues of work (BLE range from channel sounding, online experimental verification}
%\todo[inline]{\small A conclusion section is not required. Although a conclusion may review the main points of the paper, do not replicate the abstract as the conclusion. A conclusion might elaborate on the importance of the work
%or suggest applications and extensions.}
The results demonstrate the feasibility of our Bluetooth PARS-aided FGO-based estimator for GNSS-denied navigation at short ranges. Position is well-estimated and erroneous data are handled better by the FGO-based navigation system configurations compared to equivalent based on the ESKF. Rapid yaw rate is challenging, but may be mitigated by, e.g., INS aiding from a magnetometer or by mounting multiple PARS antennas on the drone. The two M-estimators tested achieve comparable performance, with Tukey resulting in higher estimation uncertainty in the first segment. 

Future work includes incorporating BLE range via channel sounding in order to make a fully-fledged Bluetooth-aided INS. Making use of a more sophisticated calibration method to determine the antenna orientation is also of particular interest, as this can make the system more suitable for use in real-time experiments down the line.

\begin{ack}
The authors would like to thank colleagues at NTNU: Johan Suárez at the Department of Electronic Systems and Pål Kvaløy and Morten Einarsve at the Department of Engineering Cybernetics for their help in conducting field experiments with the multirotor drone. Johan Suárez also participated in the post-processing and analysis work involved in converting the raw BLE PARS measurements into angular measurements.
\end{ack}

%\section{Declaration of Generative AI and AI-assisted technologies in the writing process} 
\vspace*{-0.075cm}
\bibliography{references_shortened,additional}             % bib file to produce the bibliography

\end{document}